\documentclass{article}
\usepackage{graphicx}
\usepackage{url}
\usepackage{nth}
\usepackage{pst-text}
\usepackage{siunitx}  
\usepackage{amssymb}  
\usepackage{MnSymbol}
\usepackage{amsthm}

\def\CiteSelf/{{yes}}

\ifdefined\CiteSelf
\newcommand{\mycite}[1] {\cite{my-#1}}
\else
\newcommand{\mycite}[1] {\cite{#1}}
\fi

\def\Bezier{B\'ezier }		

\title{
Ordinary Facet Angles of a Stroked Path Tessellated by Uniform Tangent Angle Steps Are Bounded by Twice the Step Angle
\vspace{1em}
\\
\normalsize
Supplementary Material for SIGGRAPH 2020 Technical Paper \\ {\em Polar Stroking: New Theory and Methods for Stroking Paths}
}
\author{Mark J. Kilgard \\ NVIDIA}
\date{\today}

\begin{document}

\maketitle

\begin{abstract}
We explain geometrically why ordinary facet angles of a stroked path tessellated from uniform tangent angle steps
are bounded by twice the step angle. This fact means---excluding a small number of extraordinary facet angles straddling offset cusps---our polar stroking method bounds the facet angle
size to less than $2 \theta$ where $\theta$ is the tangent step angle.
\end{abstract}

We assume the context of our paper \mycite{polar-stroking} on {\em polar stroking} and rely on its notation. We also
rely on the analysis of offset curves provided by Farouki and Neff~\cite{Farouki:1990:APP:87526.87543,Farouki:1990:APP:87526.87544}.

\begin{figure}
\centering
\includegraphics[width=0.85\textwidth]{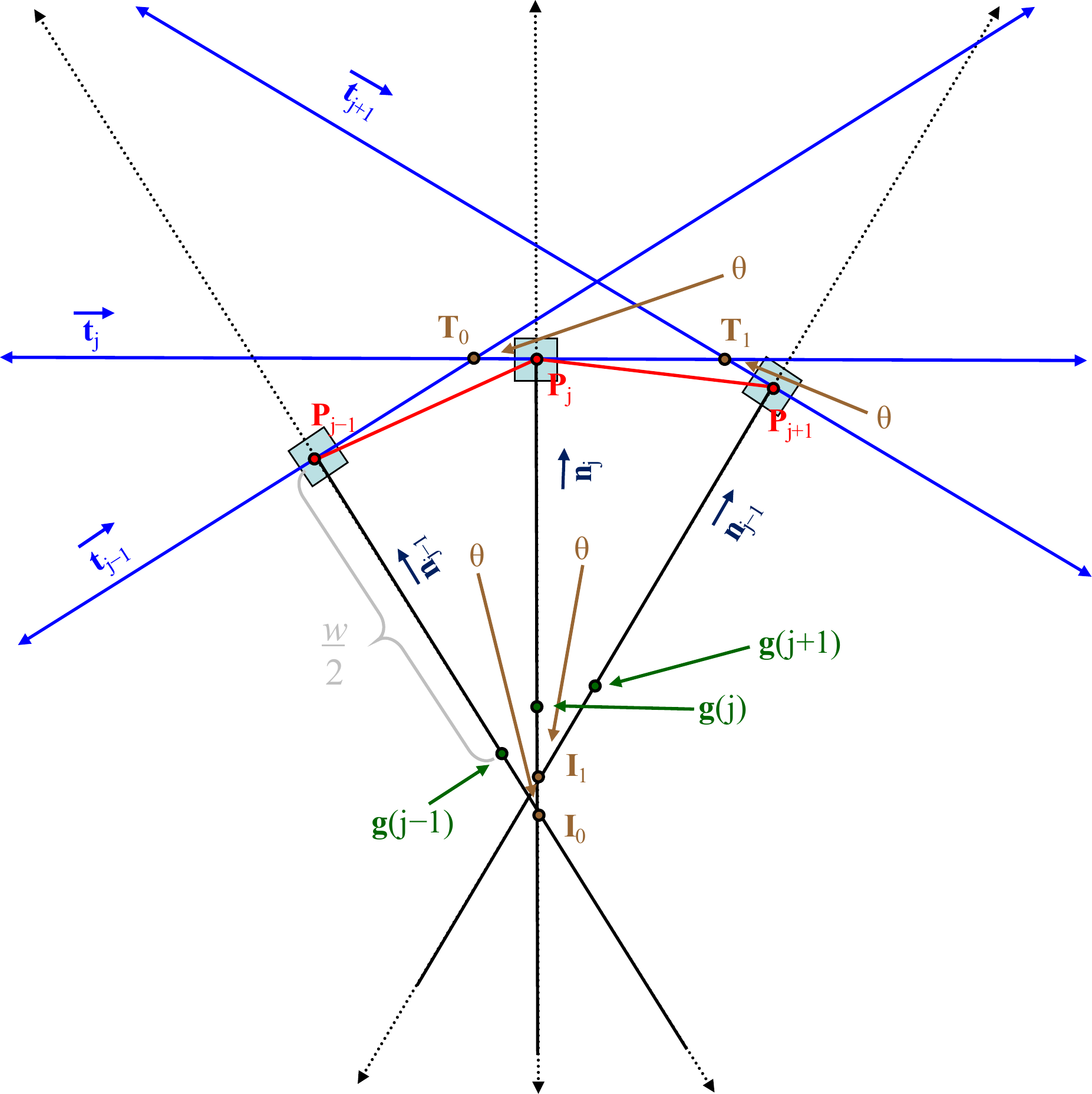}
\caption{\normalfont
Geometric configuration for a facet at the junction of two {\em red} boundary segments of a tessellated stroked path.  The facet represents any (ordinary) facet along a sequence of tessellated quads approximating a stroked path segment.
The {\em red} points $\mathbf{P}_{j-1}$, $\mathbf{P}_{j}$, and $\mathbf{P}_{j+1}$ are each a positive offset
of $\frac{w}{2}$ (half the stroke width)
extended from their respective {\em green} generator point $\mathbf{g}(j-1)$, $\mathbf{g}(j)$, and $\mathbf{g}(j+1)$ in the direction of each generator point's respective normal $\mathbf{n}_{i-1}$, $\mathbf{n}_{i}$, and $\mathbf{n}_{i+1}$.
Those generator points 
belong to the segment's generator curve $\mathbf{g}$ (not shown) such that each point in the sequence steps by a uniform tangent angle change of $\theta$ from its prior point.
\vspace{0.5em}
\newline
{\em Brown} points $\mathbf{I}_0$ and $\mathbf{I}_1$ are normal line intersections at
$\overline{\mathbf{P}_{j-1} \mathbf{g}(j-1)} \cap \overline{\mathbf{P}_{j} \mathbf{g}(j)}$
and 
$\overline{\mathbf{P}_{j} \mathbf{g}(j)} \cap \overline{\mathbf{P}_{j+1} \mathbf{g}(j+1)}$; additional brown points $\mathbf{T}_0$ and $\mathbf{T}_1$ are tangent intersections at
$\overline{\mathbf{P}_{j-1} \mathbf{t}(j-1)} \cap \overline{\mathbf{P}_{j} \mathbf{t}(j)}$
and 
$\overline{\mathbf{P}_{j} \mathbf{t}(j)} \cap \overline{\mathbf{P}_{j+1} \mathbf{t}(j+1)}$.
}
\label{fig:basic-view}
\end{figure}

Figure~\ref{fig:basic-view} shows a configuration of three successive ribs along a quadrangulation of a stroked path segment constructed using our polar stroking method.  The figure highlights a facet on the tessellation boundary.  The caption explains the configuration.

By construction, the measured internal angles formed at $\mathbf{I}_0$ and $\mathbf{I}_1$ equal the tangent angle step $\theta$.  Directed tangent lines $\mathbf{t}_{j-1}$, $\mathbf{t}_j$, $\mathbf{t}_{j+1}$ and normal lines $\mathbf{n}_{j-1}$, $\mathbf{n}_j$, $\mathbf{n}_{j+1}$ respectively meet orthogonally at $\mathbf{P}_{j-1}$, $\mathbf{P}_{j}$, and $\mathbf{P}_{j+1}$ (indicated by cyan squares in the figure).  By the {\em angle sum of a convex quadrilateral} and {\em supplementary angle} properties, the measure of internal angles formed at $\mathbf{T}_0$, and $\mathbf{T}_1$ also equal $\theta$.
The facet angle $f$ is external to $\angle \mathbf{P}_{j-1} \mathbf{P}_{j} \mathbf{P}_{j+1}$ so
\begin{align}
\measuredangle f = 180\si{\degree} - \measuredangle   \mathbf{P}_{j-1} \mathbf{P}_{j} \mathbf{P}_{j+1}
\label{eq:quad-sum}
\end{align}

\begin{figure}
\centering
\includegraphics[width=\textwidth]{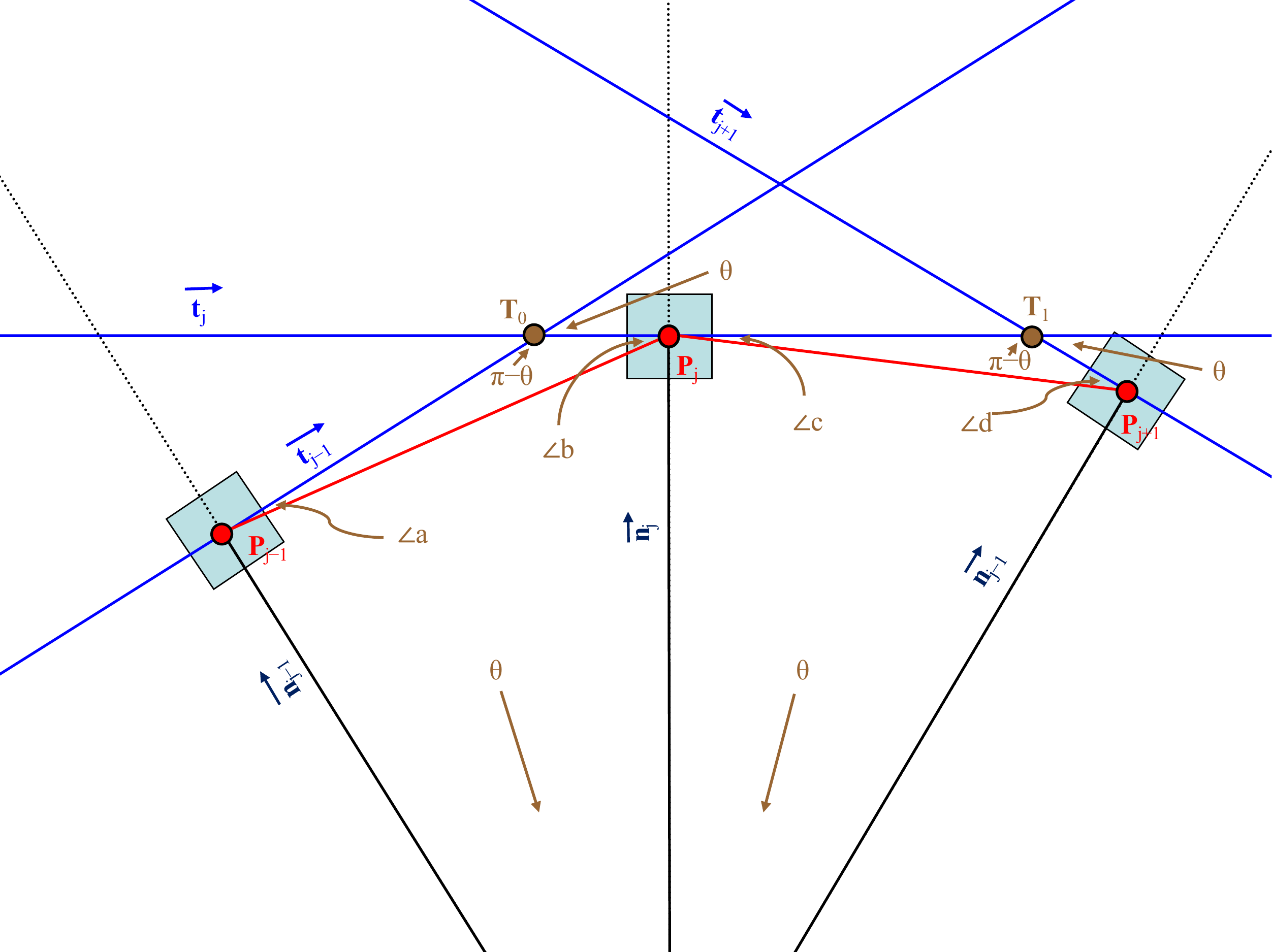}
\caption{\normalfont
Zoomed in version of Figure~\ref{fig:basic-view} to highlight the triangles bracketing the facet angle.
The facet angle $f$ measures $\measuredangle b + \measuredangle c$.
}
\label{fig:zoomed-view}
\end{figure}

\begin{figure}
\centering
\includegraphics[width=0.7\textwidth]{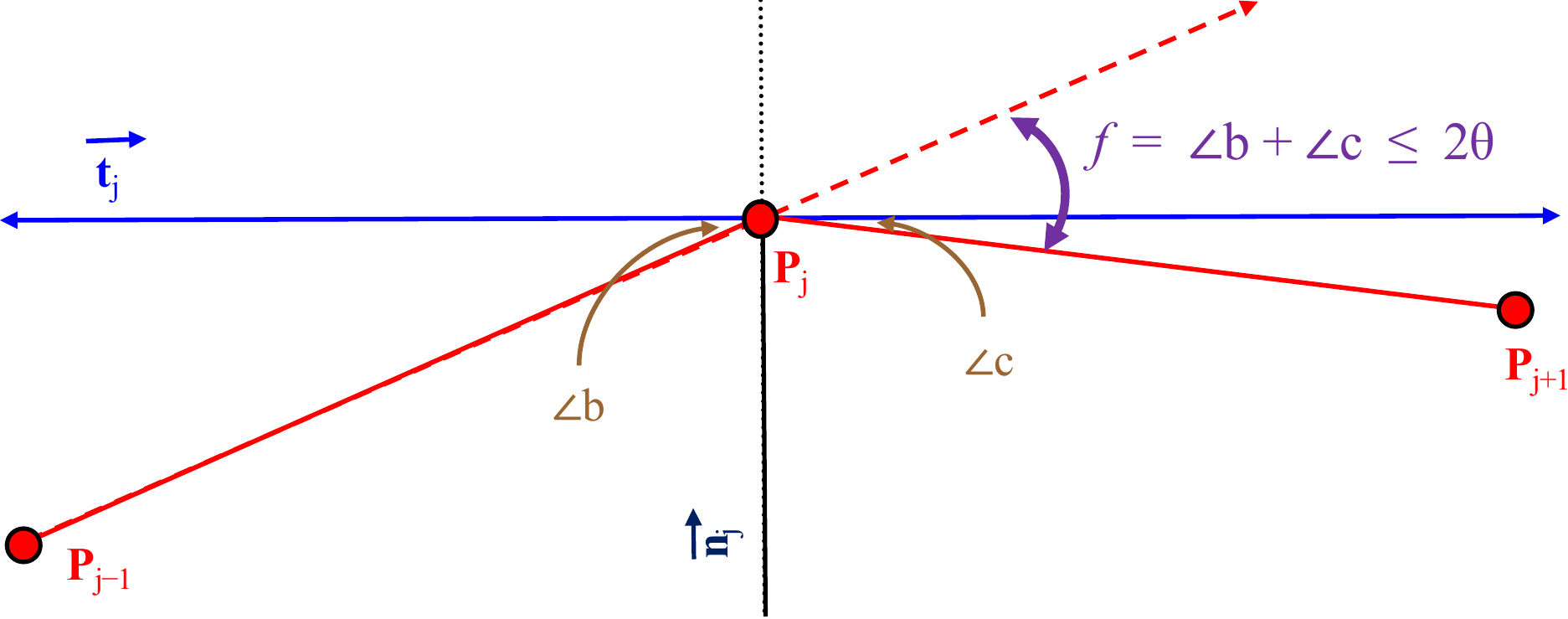}
\caption{\normalfont
Isolating just the facet angle $f$ from Figure~\ref{fig:zoomed-view}.
}
\label{fig:facet-angle}
\end{figure}

Figure~\ref{fig:zoomed-view} zooms into the facet region of Figure~\ref{fig:basic-view} and labels additional angles adjacent to the facet.
Figure~\ref{fig:facet-angle} shows the facet angle $f$ in isolation for clarity.

The tangent angle step $\theta$ relates to the tangent angle step threshold $q$ in our paper such that $\theta$ should be close-to-but-less-than $q$ while $\frac{\delta_k}{\Delta_k}$ serves as the actual $\theta$ value for a particular tangent angle interval range $k$ of our polar stroking method.

Be aware the measures of the step and facet angles shown in our figures are greatly exaggerated in order to construct expositive figures.  In the practice of polar stroking a facet angle will be small enough to not be easily discernible, likely on the scale of a few degrees or less. 

By the {\em angle sum of a triangle} property
\begin{align*}
\measuredangle a + \measuredangle b + (\pi - \theta) = \pi = 180\si{\degree}
\\
\measuredangle c + \measuredangle d + (\pi - \theta) = \pi = 180\si{\degree}
\end{align*}
So
\begin{align*}
\theta = \measuredangle a + \measuredangle b  \\
         = \measuredangle c + \measuredangle d 
\end{align*}
and when combined 
\begin{align*}
\measuredangle a + \measuredangle b + \measuredangle c + \measuredangle d = 2 \theta
\end{align*}
Allowing both $a$ and $d$ to diminish to zero means
\begin{align}
\measuredangle f = \measuredangle b + \measuredangle c \leq 2 \theta
\label{eq:bound}
\end{align}
Equation~\ref{eq:bound} bounds each so-configured facet angle $f$ to $2 \theta$ in the stroked path's quadrangulation by uniforms steps of $\theta$ in tangent angle (i.e., polar stroking).

We hedge our claim saying ``so-configured'' because, as we shall explain, not all facet angle configurations are what we call {\em ordinary} facet angles.  There may be a small finite number of
{\em extraordinary} facet angles that do not bound the facet angle.  These facet angles straddle cusps on the offset curve $\mathbf{g}_o$ where the curvature is unbounded and so cannot be expected to have a bounded facet angle.

\begin{figure}
\centering
\includegraphics[width=\textwidth]{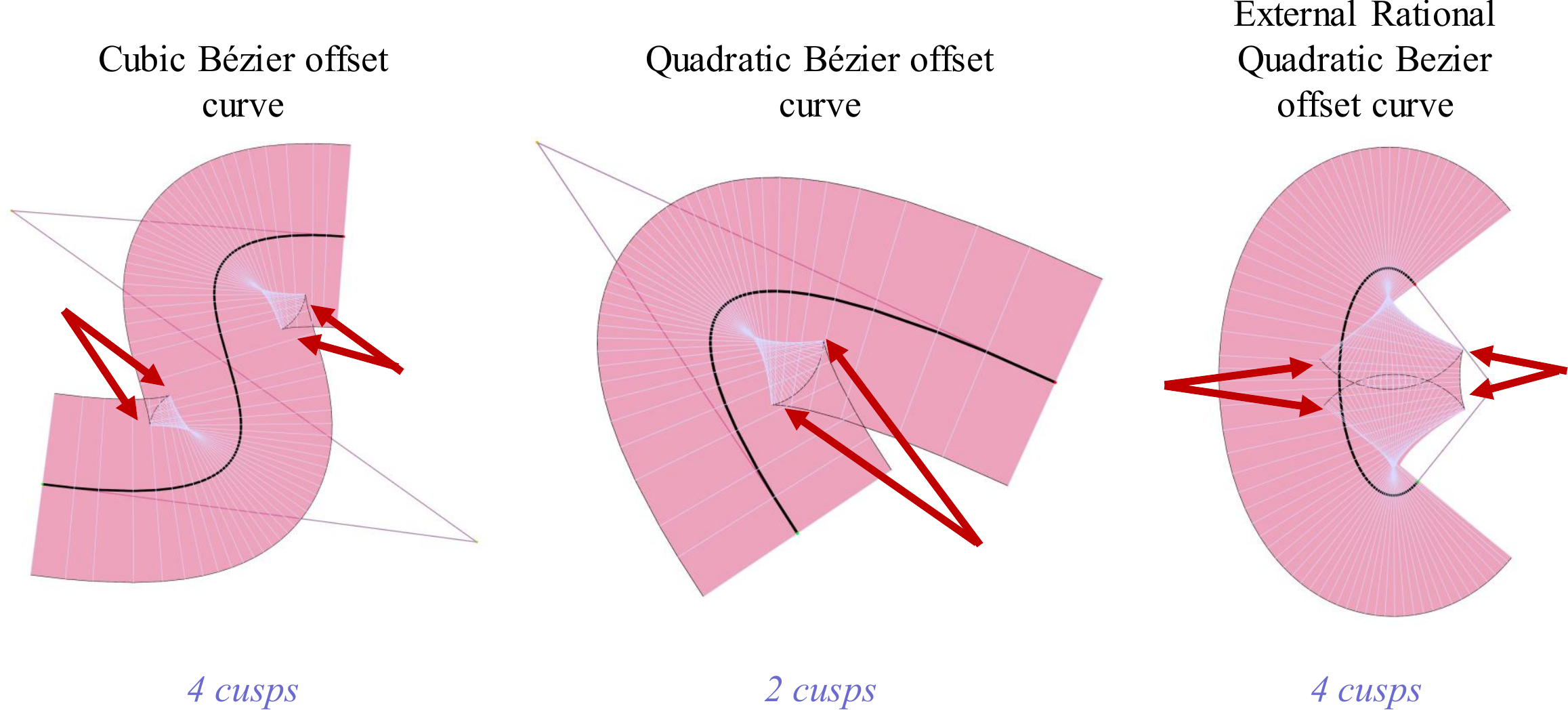}
\caption{\normalfont
Offset cusp examples on stroked path segments.  The tessellation by polar stroking uses $q=4$.  The cusps are internal to the cubic and quadratic \Bezier examples, but the rightmost example shows that cusps can be exposed too.
}
\label{fig:offset-cusp-examples}
\end{figure}

Examples of such offset cusps are shown in Figure~\ref{fig:offset-cusp-examples}.  Each cusp looks like a talon and identifies a point of the offset curve $\mathbf{g}_o$ with unbounded curvature.  Often in practice, these facet angles tend to be hidden internal to the stroked path region when rasterized with the standard {\em once-and-only-once} per pixel composting rule of path rendering standards so such cusps are rarely visible---though they can be when stroked paths are sufficiently wide or the cusps are near the ends of stroked segments.  Caps and joins on stroked paths also tend to hide these cusps.  As these offset cusps are intrinsic to the stroked path's offset curve and typically hidden, we judiciously exclude extraordinary facet angles that straddle such cusps from our claim to bound facet angles by $2 \theta$ by limiting our claim to ordinary facet angles.

\begin{figure}
\centering
\includegraphics[width=\textwidth]{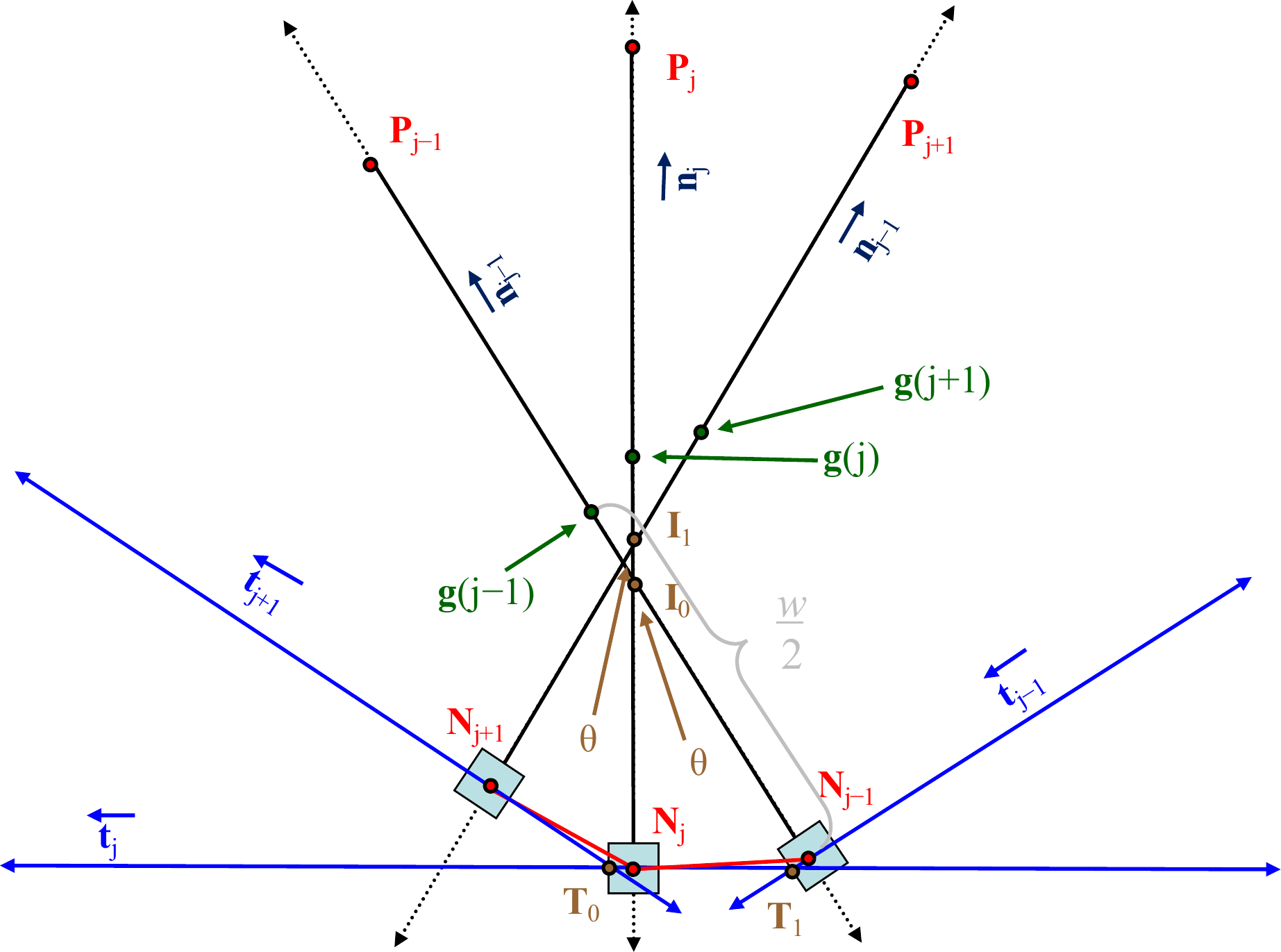}
\caption{\normalfont
Illustrate the negative tessellated boundary and its facet angle, the complement to the positive boundary shown in Figures~\ref{fig:basic-view} and \ref{fig:zoomed-view}.
The lower red points $\mathbf{N}_{j-1}$, $\mathbf{N}_{j}$, and $\mathbf{N}_{j+1}$ are a {\em negative} offset
of $\frac{w}{2}$ (half the stroke width)
from their respective generator point so opposite the direction of each generator point's respective normal $\mathbf{n}_{i-1}$, $\mathbf{n}_{i}$, and $\mathbf{n}_{i+1}$.
}
\label{fig:negative-edge}
\end{figure}

The configuration of generator points
$\mathbf{g}(j-1)$, $\mathbf{g}(j)$, and $\mathbf{g}(j+1)$ illustrated in Figures~\ref{fig:basic-view} and \ref{fig:zoomed-view} assumes
the tangent angle step winds clockwise and addresses just the positive branch of the offset curve's tessellation.
As long as generator point $\mathbf{g}(j)$ and its normal line $\mathbf{n}_j$ ``separates''' points $\mathbf{g}(j-1)$ and $\mathbf{g}(j+1)$ on opposite sides of line $\overline{\mathbf{P}_{j} \mathbf{n}(j)}$, by arguments of symmetry, the same bound applies to counterclockwise winding configurations as well as configurations involving the offset curve's negative branch as depicted in Figure~\ref{fig:negative-edge}.

The construction of Figure~\ref{fig:basic-view} only requires the normal lines $\mathbf{n}_{j-1}$, $\mathbf{n}_j$, and $\mathbf{n}_{j+1}$ to step uniformly by $\theta$.
With each step in $j$, successive generator point $\mathbf{g}_{j+1}$ identifies a next point along the generator curve $\mathbf{g}$ that steps the tangent angle smoothly by $\theta$. Each offset point $\mathbf{P}_{j+1}$ and $\mathbf{N}_{j+1}$ also advances smoothly by the same tangent angle step $\theta$ but with a different gradient magnitude because each is offset normal to $\mathbf{g}_{j+1}$ by $\frac{w}{2}$.

As long as $\mathbf{g}_{j+1}$, $\mathbf{P}_{j+1}$, and $\mathbf{N}_{j+1}$ all advance to stay in front of the line $\overline{\mathbf{P}_{j} \mathbf{n}(j)}$, our configuration stays ordinary as in Figure~\ref{fig:basic-view}.

We know from Farouki and Neff~\cite{Farouki:1990:APP:87526.87543} that the curvature of an offset curve $\mathbf{g}_o$ generated from $\mathbf{g}$ has a curvature expressible in terms of the curvature of the generator curve's curvature $\kappa_g$:
\begin{align}
\kappa_o = \frac{\kappa_g}{| 1 + \kappa_g \frac{w}{2}|}
\label{eq:offset-curve-curvature}
\end{align}
So when $1 + \kappa_g \frac{w}{2}$ passes through zero a cusp will form allowing the offset curve to ``back track'' but this 180\si{\degree} reversal happens only at cusps. As $1 + \kappa_g \frac{w}{2}$ is algebraic, we can solve for the finite number of solutions to identify a finite number of cusps.

\begin{figure}
\centering
\includegraphics[width=0.95\textwidth]{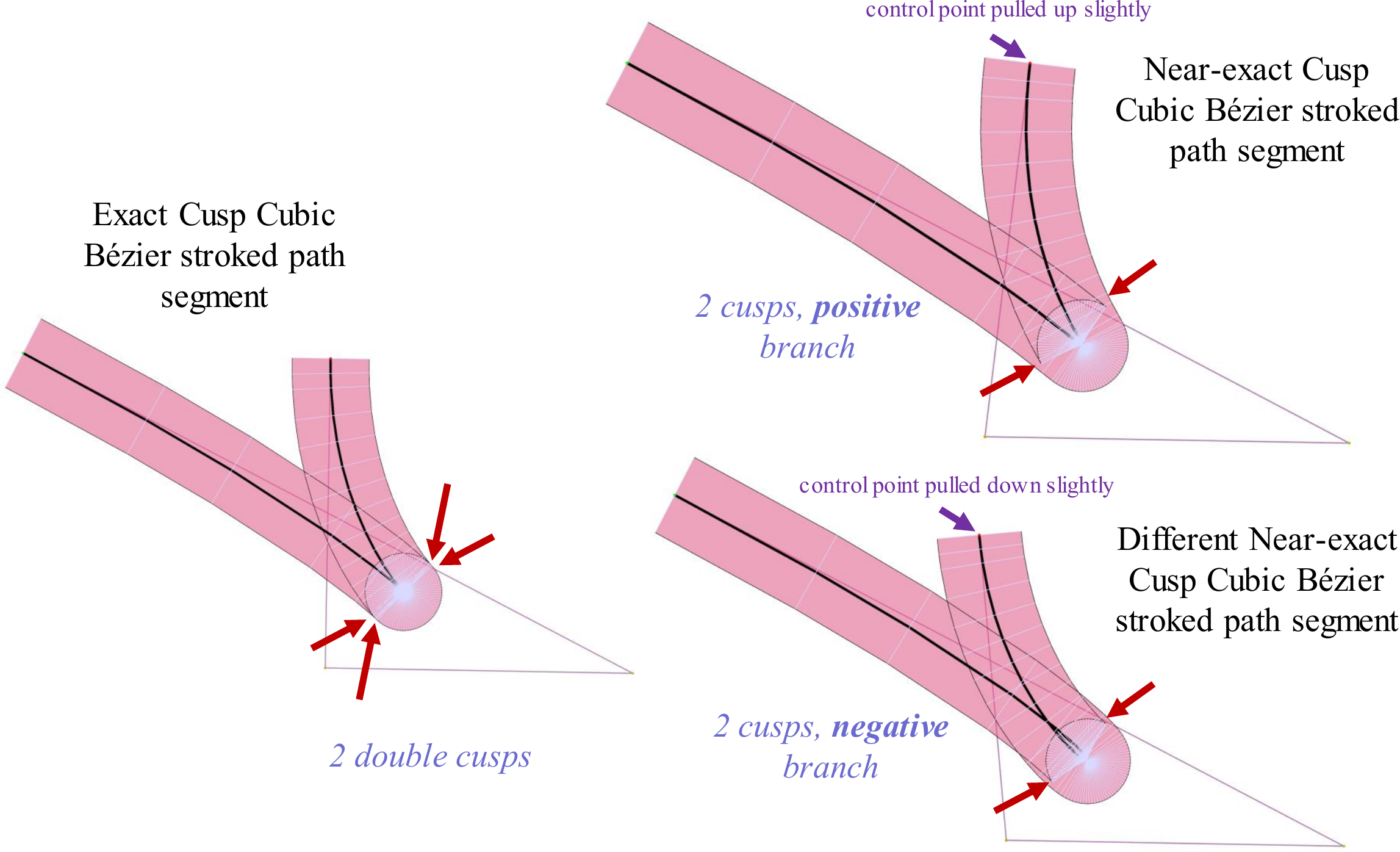}
\caption{\normalfont
Exact cusp (left) and two almost-cusp cases (right).  Notice how the cusps migrate to either the negative or positive branch of the offset curve after a slight perturbation to the red control point.
}
\label{fig:cusp-examination}
\end{figure}

We mention in passing the possibility that $\kappa_g$ itself might contain a cusp, but this is a case polar stroking already handles robustly with a tessellation that guarantees a facet angle bound $\leq q$ when polar stroking generates a double semicircle around such cusps.  In this case, all the cusp ribs intersect the cusp point and form isosceles triangles.  This would imply
\begin{align*}
\measuredangle a = \measuredangle b = \measuredangle c = \measuredangle d = \theta/2
\end{align*}
and this exact equality for the facet angle at a cusp $f_{\mbox{cusp}}$
\begin{align*}
f_{\mbox{cusp}} = \measuredangle b + \measuredangle c =\theta
\end{align*}
Figure~\ref{fig:cusp-examination} (left) shows this behavior for an exact cusp as well as portraying the tessellation of near cusps.

Now we return to consideration of offset cusps\ldots

An offset cusp allows
$P_{j+1}$ (or by symmetry any of $P_{j-1}$, $N_{j-1}$, or $N_{j+1}$ with appropriate modifications) to move such that the quadrilateral formed by $I_0$, $P_j$, $T_0$, and $P_{j-1}$ may not form a convex quadrilateral because it allows $P_{j+1}$ to ``reverse direction'' so that it and $P_j$ straddle an offset cusp.  This makes it possible for the quadrilateral to ``camel-back'' (so concave) or ``bow-tie'' (so self-intersecting) instead of being convex.  Such a non-convex quadrilateral invalidates the justification for Equation~\ref{eq:quad-sum} and hence would invalidate our claim to have a bound on the facet angle.

Such a move is only possible because the step in tangent angle ``moved through'' a cusp on the smooth offset curve $\mathbf{g}_o$ that reverses its tangent line direction.  
Otherwise---when a facet edge does not straddle a cusp on $\mathbf{g}_o$---no such reversal is possible and ordinary convex quadrilateral configurations result.

\begin{figure}
\centering
\includegraphics[width=\textwidth]{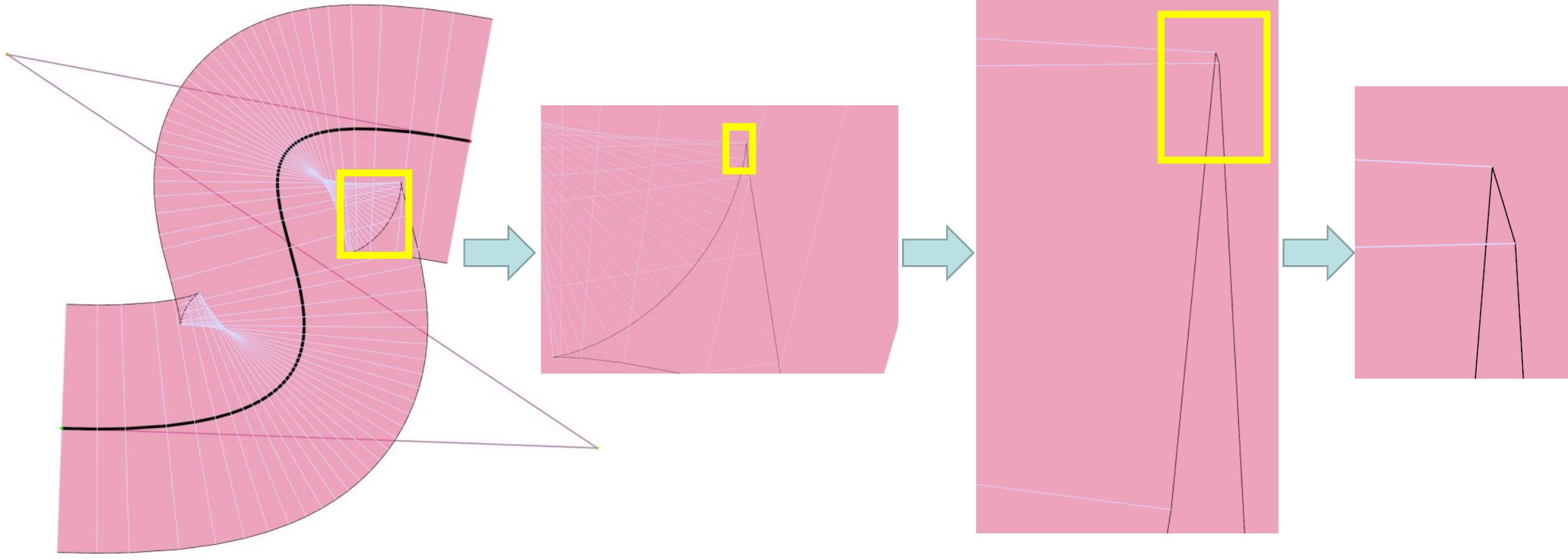}
\caption{\normalfont
Zooming into an extraordinary facet angle of a cubic \Bezier path segment tessellated with polar stroking shows an extraordinary facet angle that varies by substantially more than 8\si{\degree} (double $q=4$).
}
\label{fig:zoom-to-extraordinary-facet-angle}
\end{figure}

You can observe offset cusps in Figure~\ref{fig:offset-cusp-examples}, but the scale at which an extraordinary facet angle forms have such a minute scale that it is not observable at the resolution of the image.  But if you zoom in sufficiently you will see the problem.  For the facet angles that ``straddle'' a cusp on $g_o$, the actual facet angle becomes unpredictable and can induce facet angles approaching 90\si{\degree} in worst case situations.  Figure~\ref{fig:zoom-to-extraordinary-facet-angle} shows an example.

We briefly consider the possibility that forward steps (so not moving backward due to an offset cusp) could still induce a non-convex quadrilateral.  We require uniform steps in tangent angle and
expect these steps to be small.
But even when we conservatively assume $\theta < 90\si{\degree}$ this heads off a single step by $\theta$ being sufficient to ``turn around'' the curve by stepping forward on step.  Recall from our paper (Section 3.3.5, {\em Bounding Total Curvature Within Tangent Angle Intervals}) that every interval for the generator curves used by vector graphics standards is split to be less than 180\si{\degree}.

Geometrically, consider what happens when a facet of the tessellation of $\mathbf{g}_o$ straddles a cusp on $\mathbf{g}_o$.  This effectively ``clips off'' the cusp. Because these facet angle changes typically are internal to the stroked region, this is not objectionable for rendering and the smaller $\theta$ is, the smaller the cusp tip that is clipped away.  

This issue of a tessellated boundary segment approximating offset curve straddling an offset curve cusp is an inherent problem with quadrangulation of stroked paths modeled by offset curves.

One way around this problem is actively solving for the position of offset cusps and breaking the segment into intervals.  As a practical matter solving to exactly identify cusp offsets would be difficult as the equations of offset curves is known to be significantly higher order than the generating curve~\cite{Farouki:1990:APP:87526.87544}.  The artifacts involved are obscure so it would be more prudent to tessellate more by diminishing $\theta$.

As we show in Figure~\ref{fig:offset-cusp-examples} the worst case number of offset cusps: four for (non-rational) cubic \Bezier segments; four for rational quadratic \Bezier segments; two for nonrational quadratic \Bezier segments; and (not shown, but trivial) none for line segments.  Because a facet angle could straddle two different segments that might bound a cusp, the number of extraordinary facet angles is, in general, double the number of offset cusps.

\begin{figure}
\centering
\includegraphics[width=0.95\textwidth]{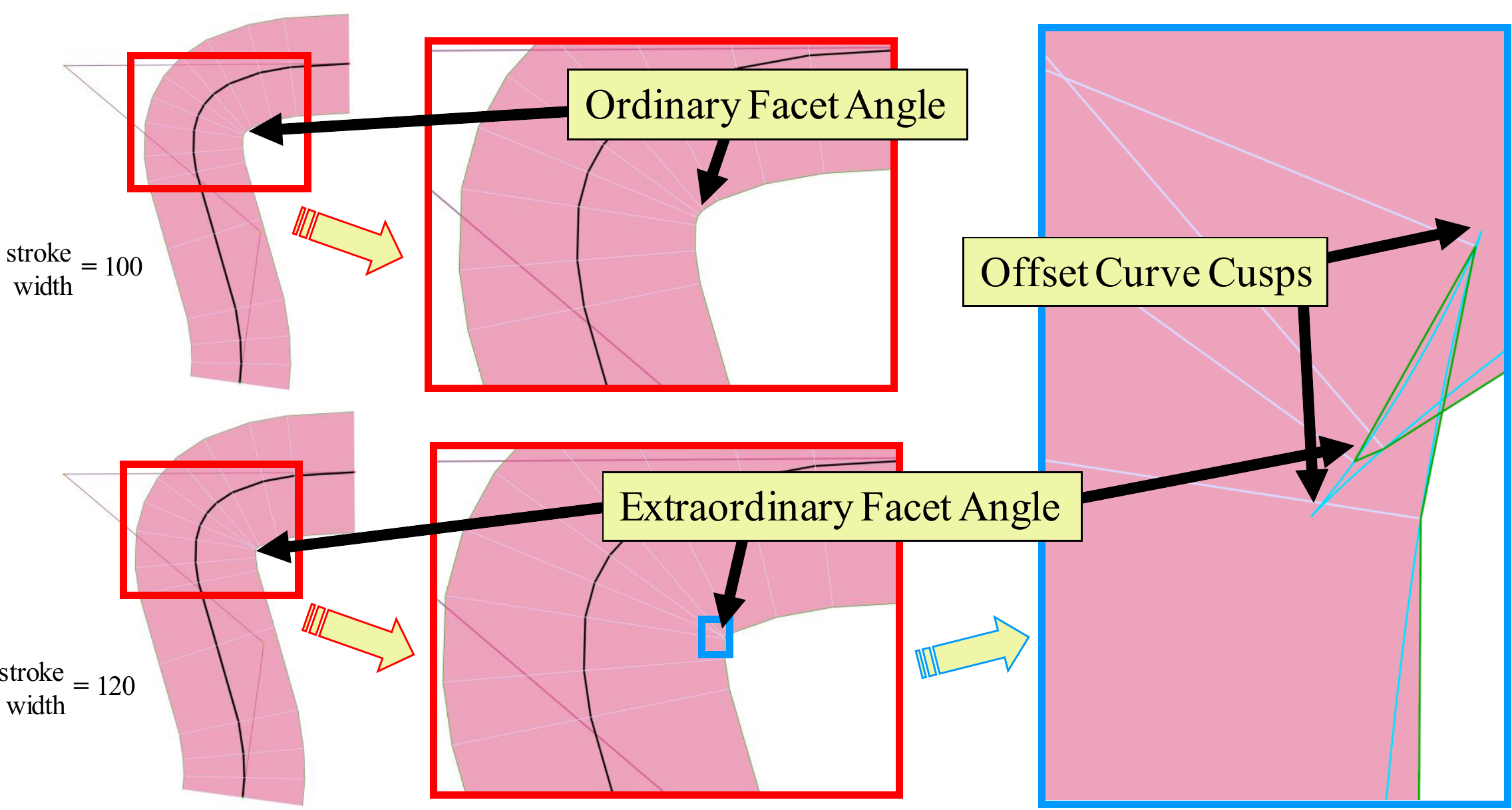}
\caption{\normalfont
Views of a cubic \Bezier segment showing how widening the stroke width from 100 (top row) to 120 (bottom row) can induce a cusp on the offset curve $\mathbf{g}_o$ that creates a step along the offset curve tessellation to ``reverse'' and create an unbounded facet angle.  The left top and bottom images show the ``zoomed out'' segment.  The middle images zooms in enough to see the region where an inner cusp will form as the stroke width increases from 100 to 120.  The right image shows the 120 stroke width version zoomed into to see the resulting extraordinary facet angle (green line segments) approximating the bright blue offset curve. The tessellation is very low quality as $q=15\si{\degree}$.
}
\label{fig:widen-to-extraordinary}
\end{figure}

To build the intuition of what happens when a cusp on the boundary of a stroked segment forms, Figure~\ref{fig:widen-to-extraordinary} shows a \Bezier segment that forms an offset curve cusp as the stroke width increases from 100 to 120.

In conclusion, we have established ordinary facet angles of a stroked path tessellated by uniform tangent angle steps are bounded by twice the step angle. This bound gives us confidence about the angular quality of stroked path tessellations generated our polar stroking method.

\bibliographystyle{plain}

\end{document}